\documentclass[aps,pre,preprint,superscriptaddress]{revtex4-1}

\usepackage{graphicx,amssymb,verbatim,amsmath,siunitx,color,float, soul}
\usepackage{natbib,url, enumerate}
\usepackage[colorlinks,allcolors = blue,bookmarksopen,bookmarksnumbered]{hyperref}
\usepackage{tabularx}
\usepackage[table]{xcolor}
\usepackage{multirow}

\usepackage{hyperref}
\usepackage{bibunits}
\defaultbibliographystyle{apsrev4-1} 
\defaultbibliography{ref}
\begin{document}
\begin{bibunit}

\title{Universal $1/f$ Noise in the Power Spectra of Energy Time-series in Solvated DNA Dynamics}

\author{Harsh Sahu}
\thanks{These authors contributed equally to this work.}
\affiliation{School of Physical Sciences, Jawaharlal Nehru University, New Delhi 110067, India}

\author{Deepika Sardana}
\thanks{These authors contributed equally to this work.}
\affiliation{School of Physical Sciences, Jawaharlal Nehru University, New Delhi 110067, India}

\author{Pramod Kumar}
\affiliation{School of Physical Sciences, Jawaharlal Nehru University, New Delhi 110067, India}

\author{Ajay Kumar Chand}
\affiliation{School of Physical Sciences, Jawaharlal Nehru University, New Delhi 110067, India}

\author{Sobhan Sen}
\email[Corresponding author:]{sens@jnu.ac.in}
\affiliation{School of Physical Sciences, Jawaharlal Nehru University, New Delhi 110067, India}

\author{Sanjay Puri}
\email[Corresponding author:]{purijnu@gmail.com}
\affiliation{School of Physical Sciences, Jawaharlal Nehru University, New Delhi 110067, India}

\begin{abstract}
We investigate energy fluctuations of solvated DNA in three conformations: G-quadruplex, Hairpin, and Duplex. Our comprehensive molecular dynamics (MD) simulations demonstrate that the time-series of interaction energies for DNA self, DNA-water and DNA-ion show $1/f$ noise with a universal decay exponent $\beta \simeq 0.90$. We show that DNA topology and structural rigidity modulate its dynamic coupling with the surrounding hydration shell and mobile counter-ions. This coupling critically shapes the $1/f$ noise profile: DNA self energies in flexible structures (G-quadruplex, Hairpin) display pronounced low-frequency modes absent in the rigid Duplex. These findings reveal how the interplay between DNA topology and environmental coupling governs different stochastic behavior in biologically relevant forms of DNA.
\end{abstract}

\maketitle

Biomolecules such as DNA, proteins, and lipids exhibit a rich spectrum of motion that spans multiple length and time scales. This motion occurs in a complex cellular environment, which is an aqueous medium crowded with ions and organelles. The dynamics of these biomolecules is closely related to their structural organization and biological functions \cite{henzler2007dynamic, bahar2010global}. A common approach to characterizing this dynamics involves the analysis of correlation functions and power spectra, which often reveal signatures of $1/f$ noise \cite{Li1992, Yang2003, Min2005, Yamamoto2021, Yamamoto2014, Bizzarri2013, Bezrukov2000, Mercik2001, Tasserit2010}. Long-range correlated dynamics is increasingly recognized as fundamental to a variety of biological processes. This is supported by observations of $1/f$ noise in various systems, including DNA sequence organization \cite{Voss1992,Li1992}, protein conformational fluctuations \cite{Yang2003,Min2005,Yamamoto2014,Yamamoto2021}, bio-recognition mechanisms \cite{Bizzarri2013}, and ionic current fluctuations through membrane channels \cite{Siwy2002, Bezrukov2000, Mercik2001, Tasserit2010}. 

\begin{figure}[htb]
\centering
\includegraphics[width=0.9\textwidth]{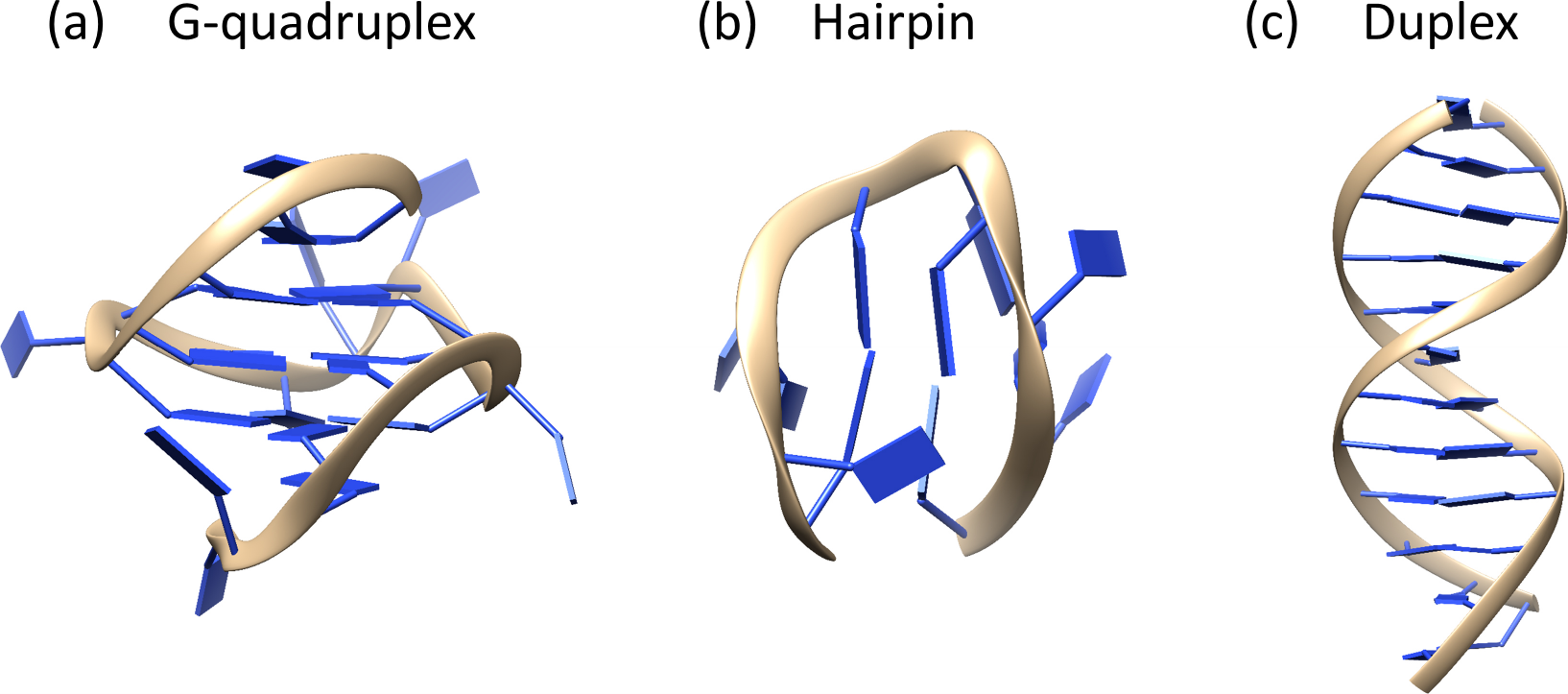} 
\caption{Schematic illustration depicting experimentally determined structures of DNA, (a) G-quadruplex (PDB id 1XAV), (b) Hairpin (PDB id 5M1W), (c) Duplex (PDB id 1D30).}
\label{fig1}
\end{figure}

DNA, often referred to as the ``molecule of life'', exists within cells in a variety of biologically relevant structural conformations. These include the canonical Watson-Crick Duplex, Hairpin structures, and higher-order G-quadruplex DNA formations (G4 DNA) (Fig.~\ref{fig1}). These diverse DNA conformations, in conjunction with water and ions, play an essential role in biological processes \cite{tan2008salt,sata2025,miller2010hydration,obara2024insights, bhattacharyya2016metal}. In this letter, we report results from comprehensive atomistic molecular dynamics (MD) simulations of these DNA structures in an environment of water and ions. Remarkably, despite their structural diversity, these DNA conformations exhibit universal features in the electrostatic interaction energies with self, surrounding water, and cations. These are particularly evident in the correlated dynamics in the long time (low frequency) regime.

Traditionally, biomolecule functions are attributed primarily to their structures, while water and ions are viewed as passive background components of the cellular environment \cite{Ball2008,Gerstein1998}. However, accumulating evidence suggests the active and dynamic roles of water molecules and ions (particularly those within the hydration shells surrounding biomolecules) in maintaining structural stability and allowing a wide range of biological functions \cite{Bagchi2013,Ball2017}. These include enzyme catalysis \cite{Grossman2011,Pal2016}, protein folding \cite{Kim2008}, DNA charge transfer \cite{ONeill2004}, drug binding to nucleic acids \cite{Mukherjee2008,Wilhelm2012}, and protein--DNA interactions \cite{Mukherjee2008,Yang2014, Sen2005}.

Temporal fluctuations with {\it power spectral density} decaying as $1/f^\beta$, where $\beta$ typically ranges from 0 to 2, are commonly referred to as $1/f$ noise \cite{btw87,mw88}. This behavior is physically significant because it characterizes correlations which decay as a power-law in time. This should be contrasted with the more generic exponential decay, which has a time scale beyond which correlations are negligible. The physical and biological implications of power-law vs. exponential decay are far-reaching and dramatic. The phenomenon of $1/f$ noise is ubiquitous and is observed in both natural and engineered systems. It was first experimentally identified in current fluctuations in thermionic vacuum tubes in the early twentieth century \cite{PhysRev.26.71}. Subsequently, $1/f$ noise has been reported in various contexts, including neuronal activity \cite{Pettersen2014}, quantum chaotic systems \cite{Riser2017,Faleiro2004}, music \cite{Voss1978}, and fluid dynamics \cite{Thompson2012,Wang2008}. Its widespread presence in different systems suggests that no single universal mechanism can explain its origin, highlighting the intrinsic complexity of $1/f$ dynamics \cite{Yadav2017,Siwy2002,Sasai1992,Voss1992}. This diversity also implies that, while the spectral signature may be similar, the physical interpretations of $1/f$ noise are often system-specific and closely related to the interactions at play. In this letter, we highlight the universal nature of $1/f$ noise in solvated DNA.

Several earlier studies have sought to understand conformational and energetic fluctuations in biological systems with aqueous environments. Although $1/f$ noise has been well characterized in proteins and lipids \cite{Yamamoto2021, Bizzarri2002, Bizzarri1997, Mukherjee2019,Mukherjee2022, Song2025, Dewey1992, Yama2015, Yamamoto2015}, its presence in DNA has not been explored much, especially through numerical simulations \cite{Voss1992,Li1992,Li2005, Mukherjee2023}. There have also been studies of $1/f$ noise in non-dynamical contexts, e.g., DNA sequences from the human blood coagulation factor VII gene \cite{Li1992}, as well as the complete set of 22 autosomes and two sex chromosomes \cite{Li2005}. These analyzes revealed that the spatial distribution of nucleotide pairs (e.g., A+G or A+T) often follows a power spectrum with a scaling analogous to $1/f$ in the low-``frequency'' regime. Notably, long-range correlations are especially prominent in non-coding regions of DNA, such as introns. The power spectra of these sequences can be approximated by $1/f^\beta$, with $\beta$ ranging from 0.5-0.85. These long-range spatial correlations are attributed to the presence of repetitive sequences in DNA.

Experimental studies on DNA have revealed anomalously slow power-law relaxation in the solvation energy dynamics of duplex and quadruplex DNA, indicating a complex energy landscape and persistent coupling between DNA and its solvent environment \cite{Somoza2004,Pal2010,Sardana2023,Verma2012,Sen2009, Andreatta2005,Sen2006,Andreatta2006,Pal2015}. This behavior suggests fundamentally different modes of interaction in DNA that must be identified. Clearly, it is essential to have comprehensive simulations of DNA dynamics to complement the experiments. Unfortunately, no detailed studies are available in the literature on the power spectral decay of different DNA structures. There is a preliminary simulation study by Mukherjee et al. \cite{Mukherjee2023}, where energy fluctuations in Single-stranded and Duplex DNA have been found to show bimodal $1/f$ behavior. However, these simulations were not performed on experimental DNA (PDB) structures. In addition, the systems did not have physiological salt concentrations. Our letter addresses this important lacuna in the literature by presenting comprehensive MD studies of the three DNA conformations shown in Fig.~\ref{fig1}: G-quadruplex (PDB id 1XAV), Hairpin (PDB id 5M1W) and Duplex (PDB id 1D30). In contrast to the earlier report of bimodal decay \cite{Mukherjee2023}, we observe universal $1/f$ noise in power spectra with a decay exponent $\simeq 0.90$ in all three systems.

We employ multi-microsecond atomistic MD simulations to systematically investigate energy power spectra of various DNA structures in the presence of water and physiological ionic strength. The details of the MD simulation are provided in the Supplementary Material (SM). We will demonstrate that this dynamics exhibits $1/f$ noise with a robust exponent. A major objective is to understand how the structural topology of DNA, particularly its intrinsic rigidity or flexibility, influences its dynamical coupling with water and counter-ions.

The fluctuations in the total electrostatic interaction energy of a DNA solution [$\delta E(t) = E(t) - \left\langle E(t)\right\rangle$] can be decomposed into contributions arising from individual components of the system, namely DNA, water, and counter-ions, as well as their cross-interaction terms:
\begin{equation} \label{electro} 
\delta E \left(t\right)=\sum_{\alpha, \beta } \delta E_{\alpha \beta}(t) .
\end{equation} 
In Eq.~(\ref{electro}), the term $\delta E_{\alpha \beta}(t)$ denotes the interaction between species $\alpha$ and $\beta$, with $\alpha= d$ (DNA), $w$ (water) or $i$ (counter-ions: Na$^+$ or K$^+$). The diagonal terms $\delta E_{dd}$, $\delta E_{ww}$, $\delta E_{ii}$ refer to the self-interaction energies of the corresponding species. As \textit{$\delta E_{\alpha  \beta}$} is symmetric under {$\alpha \leftrightarrow \beta$}, there are three cross-interaction terms: $\delta E_{dw}$, $\delta E_{wi}$, $\delta E_{di}$. Our interest is in DNA dynamics, so we focus on $\delta E_{dd}$, $\delta E_{dw}$, $\delta E_{di}$, which we designate as $\delta E_1, \delta E_2, \delta E_3$, respectively. The normalized time-correlation functions $C_{ij} (t)$ are then calculated as follows:
\begin{equation}
C_{ij} (t) = \frac{\left\langle \delta E_i (\tau) \delta E_j (\tau +t) \right\rangle}{\left\langle \delta E_i (\tau) \delta E_j (\tau) \right\rangle} .
\label{auto}
\end{equation}

\begin{figure}[htb]
\centering
\includegraphics[width=0.95\textwidth]{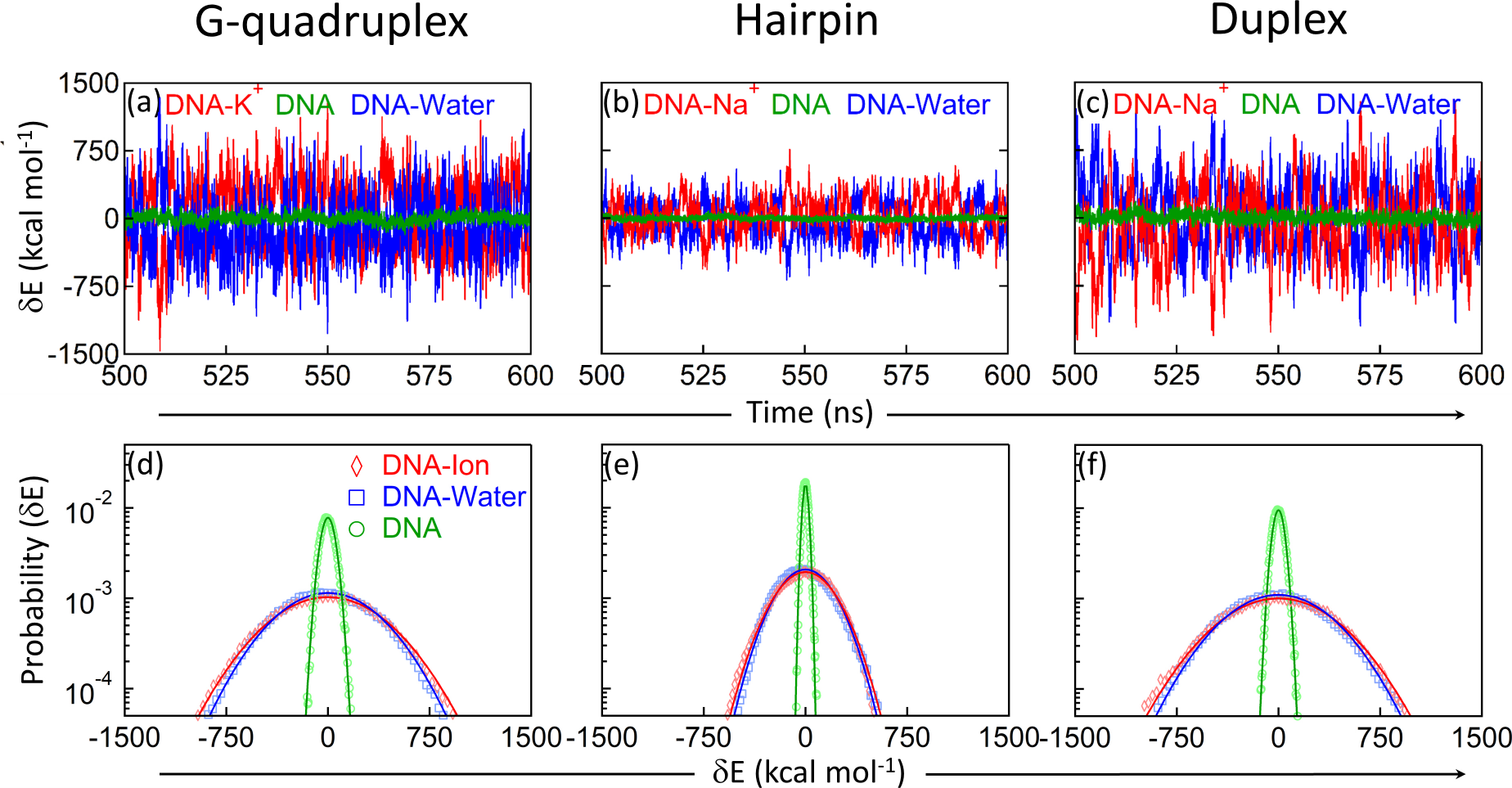} 
\caption{(a)-(c) Energy fluctuations of DNA self ($\delta E_1$, green), DNA-water ($\delta E_2$, blue), DNA-ion ($\delta E_3$, red) for all three DNA structures. Data is shown for a 100 ns window. (d)-(f) Probability distributions of energy fluctuations for all three DNA structures, calculated over the full 1000 ns trajectory. The solid lines show Gaussian fits with standard deviation $\sigma$ obtained from the data.}
\label{fig2}
\end{figure}

{\it Time Series Analysis:} Analysis of energy fluctuations from the MD trajectories of the three DNA structures revealed distinct characteristics among the components of the system. In Fig.~\ref{fig2}(a)-(c), we show typical time-series for $\delta E_1 \rightarrow \delta E_3$. The corresponding probability distributions of these quantities are shown in Fig.~\ref{fig2}(d)-(f). The probability distributions of $\delta E_1$ show sharp peaks with minimal fluctuations, while the corresponding distributions for $\delta E_2, \delta E_3$ are wider. The distributions are Gaussian in nature -- the corresponding functional forms are plotted as solid lines in Fig.~\ref{fig2}(d)-(f). [In contrast, the water self-energies (not shown here) exhibit a markedly broad distribution, reflecting the wide range of interaction strengths and orientations arising from dynamic hydrogen bonding, dipole-dipole interactions, and other intermolecular forces among water molecules.] Moreover, a robust and consistent anti-correlation was observed between DNA-water and DNA-ion energy fluctuations across all three DNA structures. This pronounced inverse relationship points to a tightly coupled interplay within the DNA hydration shell, wherein water molecules and counter-ions stochastically interchange positions.

\begin{figure}[htb]
\centering
\includegraphics[width=0.9\textwidth]{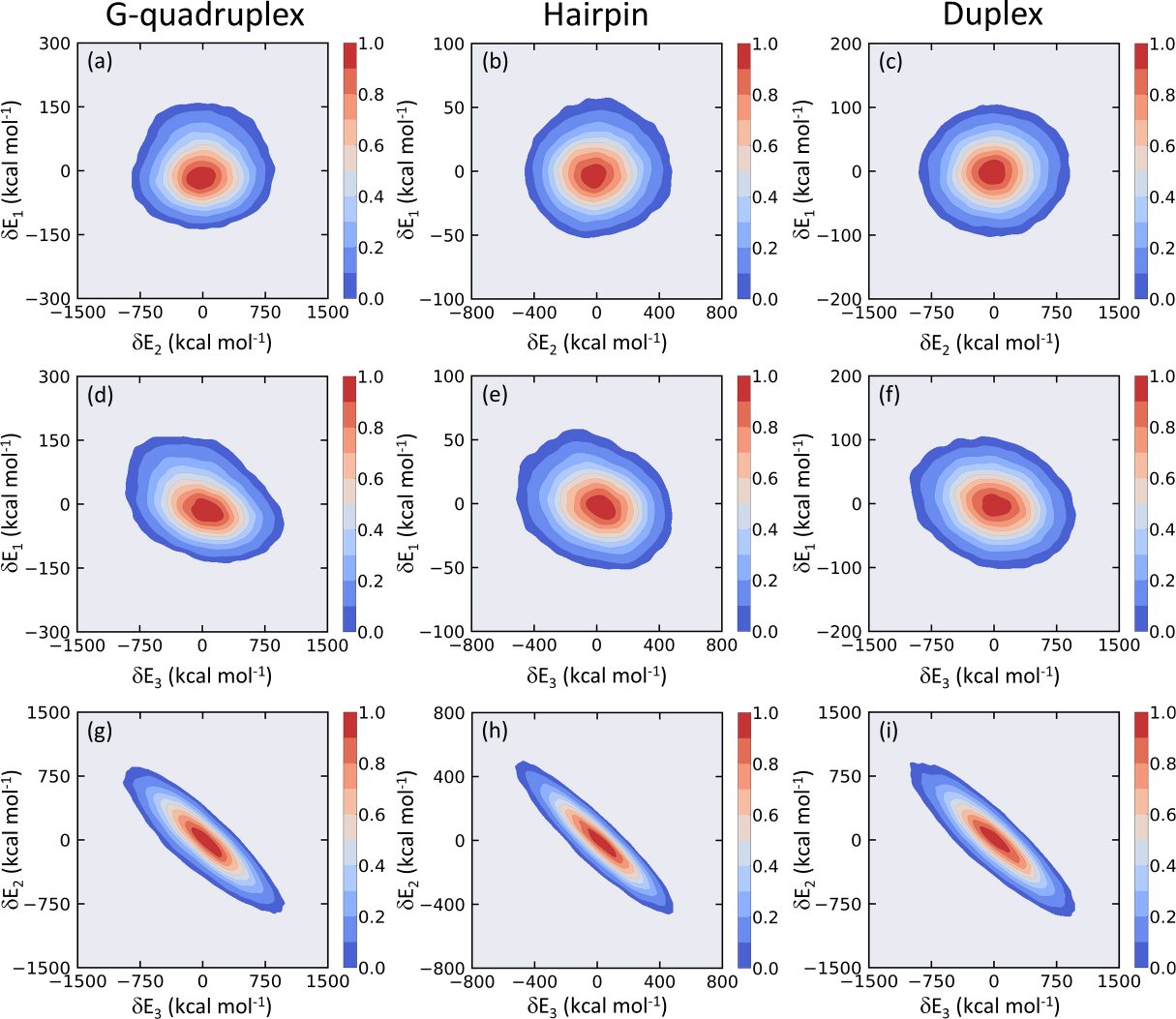}
\caption{Contour representation of bivariate distributions of $\delta E_1$, $\delta E_2$, $\delta E_3$ for (a), (d), (g) G4 DNA; (b), (e), (h) Hairpin
DNA; (c), (f), (i) Duplex DNA.}
\label{fig3}
\end{figure}

To quantify these anti-correlations, Fig.~\ref{fig3} shows contour plots of the bivariate distributions of $(\delta E_1, \delta E_2)$, $(\delta E_1, \delta E_3)$ and $(\delta E_2, \delta E_3)$. We also computed the Pearson correlation coefficient $\rho_{ij}$ for the energies:
\begin{equation}
\label{pearson} 
\rho_{ij}=\frac{\mbox{cov} (i,j)}{\sigma_i \sigma_j} .
\end{equation} 
Here, cov$(i,j)$ denotes the covariance between $i$ and $j$, while $\sigma_i$ and $\sigma_j$ are their standard deviations. Table~\ref{tab1} presents the Pearson correlation coefficients for the three DNA structures. Consider the correlation between the DNA-water and DNA-ion energies. Consistently high negative values show that when the interaction energy of DNA with water ($\delta E_2$) becomes more favorable, the interaction energy between DNA and ions ($\delta E_3$) correspondingly weakens. This can be seen in Fig.~\ref{fig3}(g)-(i), where the high probability regions correspond to anti-correlated values of $\delta E_2$ and $\delta E_3$, showing that cations and water in the vicinity of DNA frequently exchange positions and share overlapping spatial regions over time. The competition and compensation within the DNA hydration shell play a pivotal role in preserving the local electrostatic and hydration balance, thereby influencing the molecule's structural stability, dynamic behavior, and ultimately serving as a dynamic regulator of its biological function. 
\begin{table}
\centering
\begin{tabular}{|l|c|c|c|}
\hline
{\bf Interaction Energy} & {\bf G4 DNA} & {\bf Hairpin} & {\bf Duplex} \\
\hline 
DNA self ($\delta E_1)$ vs. DNA-water ($\delta E_2)$ & 0.015 & 0.029 &  0.018 \\ 
\hline
DNA self ($\delta E_1)$ vs. DNA-ion ($\delta E_3)$ & -0.320 & -0.210 & -0.200 \\
\hline
DNA-water ($\delta E_2)$ vs. DNA-ion ($\delta E_3)$ & -0.870 & -0.920 & -0.880 \\
\hline
\end{tabular}
\caption{Pearson coefficients $\rho_{ij}$ for electrostatic interaction energies across DNA structures.}
\label{tab1}
\end{table}

In contrast, Fig.~\ref{fig3}(d)-(f) shows a moderate anti-correlation between the DNA self-energy ($\delta E_1$) and DNA-ion energy ($\delta E_3$), with Pearson coefficients ranging from $-0.20$ to $-0.32$. This reflects a measurable, though less pronounced, inverse relationship. In addition, only a weak correlation is detected between $\delta E_1$ and $\delta E_2$, with very low positive $\rho_{12}$ 0.015 - 0.029. Together, these results highlight the nuanced nature of DNA self-interactions relative to DNA interactions with water and ions, underscoring the intricate dynamical interplay that governs DNA's local environment.

\begin{figure}[htb]
\includegraphics[width=0.9\textwidth]{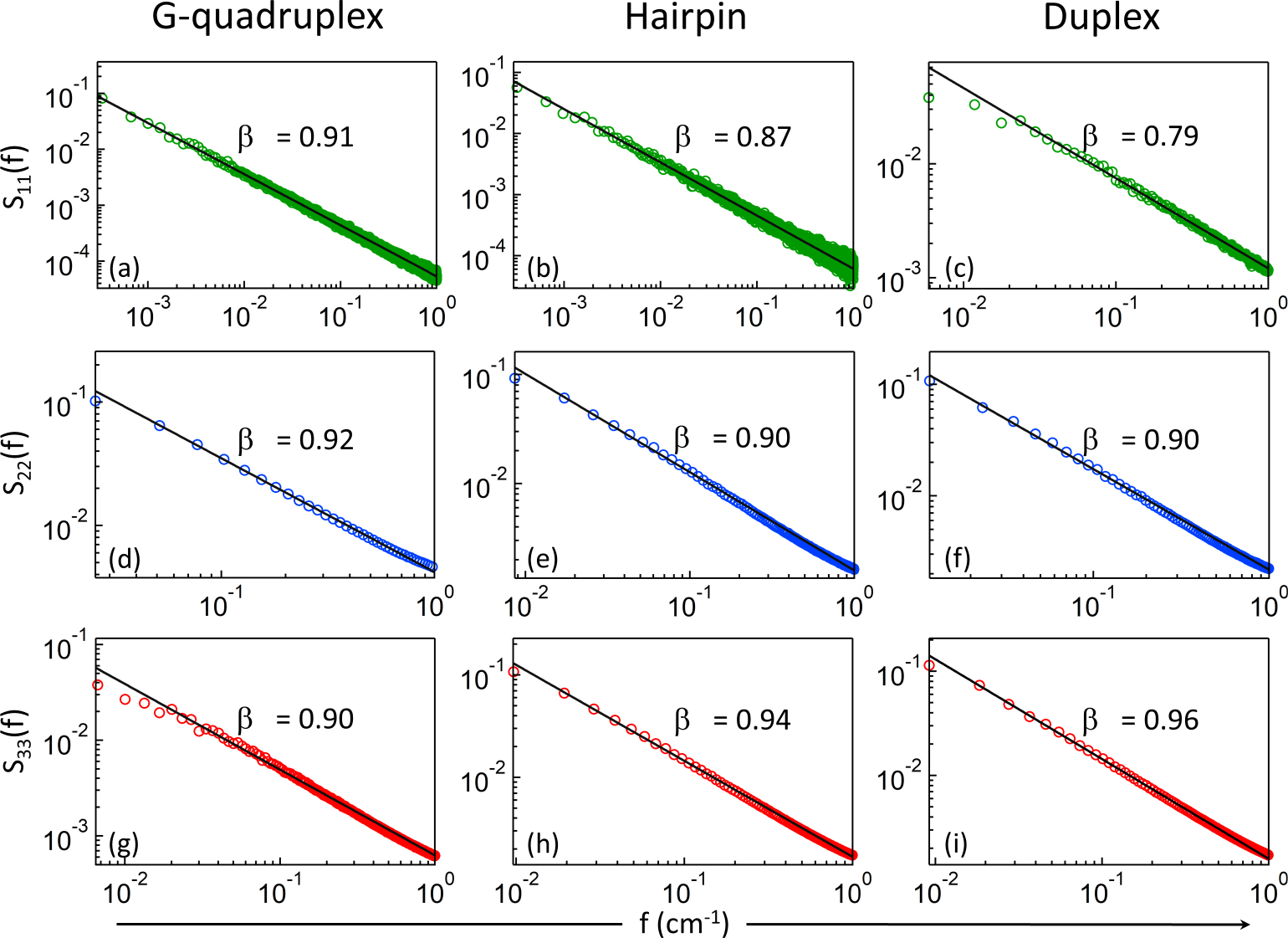}
\caption{Log-log plot of power spectra, $S_{ii}(f)$ vs. $f$, of energy time-series for $\delta E_1$, $\delta E_2$, $\delta E_3$. We show data for (a), (d), (g) G4 DNA; (b), (e), (h) Hairpin DNA; (c), (f), (i) Duplex DNA. The solid lines show best linear fits with the power-law exponent $\beta$.}
\label{fig4}
\end{figure}

{\it Power Spectra:} To elucidate the dynamical properties, we examined the autocorrelation functions $C_{ii}(t)$ and the power spectra $S_{ii}(f)$ for the energy components in the three DNA structures. These quantities are related by a Fourier transform, so we study one of them, namely $S_{ii}(f)$. (Please see Sec.~B of the SM for numerical details about how $S_{ii}(f)$ is computed.) In Fig.~\ref{fig4}, we plot $S_{11}(f)$, $S_{22}(f)$ and $S_{33}(f)$ for the three DNA structures. Strong universality is seen in these power spectra: they show distinct $1/f$ noise at moderate and higher frequencies. The solid lines denote the best linear fits to the numerical data on a log-log scale. The best-fit exponents $\beta$ are specified in each frame -- the unimodal $1/f$ behavior is seen across multiple decades and has a universal exponent $\beta \simeq 0.90$. This demonstrates a universal dynamic signature that governs the coupling between DNA and its surrounding solvent and the ionic environment, regardless of the specific structural topology of the DNA.

In physical systems, a power spectrum $S(f) \propto 1/f^\beta$ (typically $\beta \approx 1$) arises from the superposition of many independent relaxation processes with a broad distribution of relaxation times $\tau_0$:
\begin{equation}
S(f) \sim \int_0^\infty \frac{\tau_0}{1 + (f \tau_0)^2}~P(\tau_0)~d\tau_0 .
\label{dist}
\end{equation}
The kernel in Eq.~(\ref{dist}) arises from the Fourier transform of an autocorrelation function with a single exponential relaxation of the time-scale $\tau_0$. If $P(\tau_0) \sim 1/\tau_0^\theta$, the resulting power spectrum scales with frequency as $S(f) \sim 1/f^{2-\theta}$. In solvated DNA, the conjunction of multiple physical mechanisms produces these broadly distributed relaxation scales, namely counterion dynamics, DNA conformational dynamics, coupled hydration shell and hydrogen-bond networks, hierarchical energy landscapes, etc.

We highlight some important features of the power spectra of the self-interaction energy of DNA $\delta E_1$. In both the G4 DNA and Hairpin structures, this component shows pronounced low frequency modes, and a single power-law scaling extending over almost four decades. This feature is not as pronounced in Duplex DNA [Fig.~\ref{fig4}(c)], where the power law is confined to approximately 2 decades in the frequency range. (This is also true for $S_{22}(f)$ and $S_{33}(f)$ for all DNA.) This is due to the presence of long-time relaxation modes in the G4 DNA and Hairpin structures. Structural flexibility is a primary driver of long-range correlations in DNA dynamics. The larger-amplitude conformational fluctuations of G-quadruplex and Hairpin structures, arising from their reduced rigidity and flexible loop regions, facilitate the emergence of long-time dynamics in their internal energy fluctuations, manifesting as prominent low-frequency modes and extended power-law scaling in their power spectra. A spectral analysis of $S_{11}(f)$ for duplex DNA over an extended frequency range (corresponding to 1000 ns of time correlation) shows white noise in the low frequency regime, in contrast to the bimodal behavior reported in Ref.~\cite{Mukherjee2023}. This shows the absence of long-time correlations of structural fluctuations in the rigid duplex DNA (see Fig.~1 in SM). Such resilient stability of the Duplex DNA structure is produced by the physiological ionic strength (100 mM NaCl).

\begin{figure}[htb]
\includegraphics[width=1.0\textwidth]{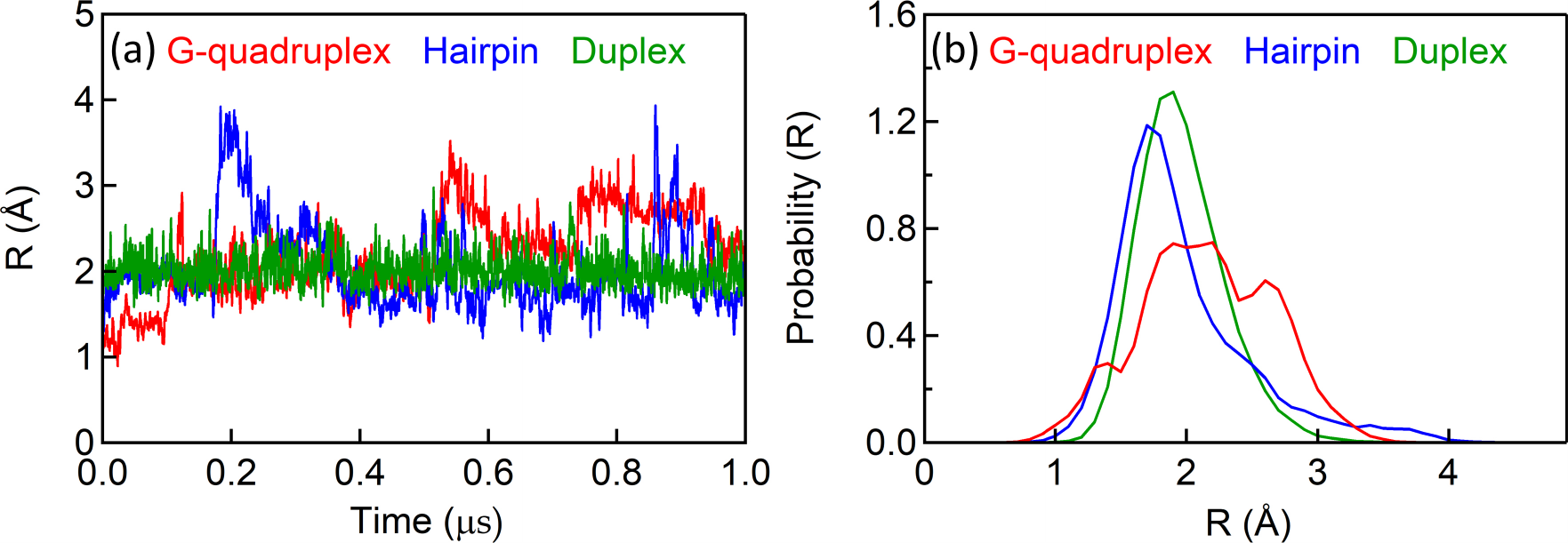}
\caption{(a) Time evolution of the root-mean-square deviation $R(t)$ for the DNA backbone of G4 DNA (red), Hairpin DNA (blue), Duplex DNA (green). (b) Probability distribution of $R$ for the three systems.}
\label{fig5}
\end{figure}

The proposed link between increased structural fluctuations in G4 DNA and Hairpin DNA and the emergence of low-frequency modes is strongly supported by studying the root mean square deviation (RMSD). This is defined as
\begin{equation}
R(t) = \left\{ \frac{1}{N} \sum_{i=1}^{N} \left[ (x_i(t) - x_i(0))^2 + (y_i(t) - y_i(0))^2 + (z_i(t) - z_i(0))^2 \right] \right\}^{1/2} .
\end{equation}
Here, $N$ is the total number of atoms in a DNA molecule; and ($x_i(t), y_i(t), z_i(t)$) are the Cartesian coordinates of the atom $i$ at time $t$. In Fig.~\ref{fig5}, we observe that G4 DNA and Hairpin DNA exhibit substantial fluctuations of RMSD that range between 1.5 Å and 4 Å during the simulation of 1 $\mu$s. In contrast, the Duplex DNA structure maintains RMSD fluctuations around 2 Å in the same time window. Thus, there is less structural variation in Duplex DNA compared to G4 DNA and Hairpin DNA. In the latter biomolecules, larger RMSD fluctuations arise from the flexible loop regions within these two structures. These conformations are intrinsic properties of the architectures, directly reflected in their extended power-law scaling of $1/f$ noise over a broad frequency range.

\begin{figure}[htb]
\includegraphics[width=0.8\textwidth]{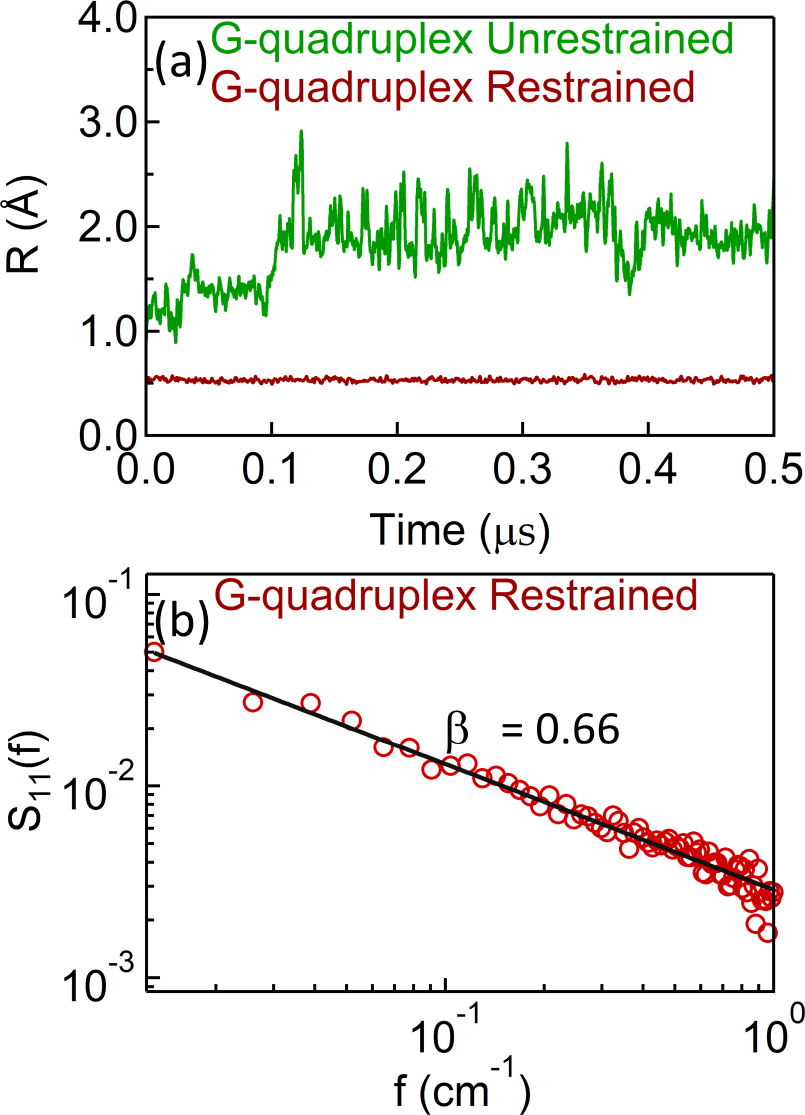}
\caption{(a) Time evolution of $R(t)$ for the DNA backbone of Unrestrained G4 DNA (green), Restrained G4 DNA (red). (b) Log-log plot of $S_{11}(f)$ vs. $f$ for Restrained G4 DNA. The solid line shows the best power-law fit with exponent $\beta$.}
\label{fig6}
\end{figure}

To directly test the link between low structural rigidity and the origin of low-frequency modes, we performed an additional 500 ns simulation in which G4 DNA atoms were restrained using a harmonic potential of 0.1 kcal mol$^{-1}$ {\AA}$^{-2}$. Restrained G4 DNA showed markedly reduced fluctuations in RMSD (see Fig.~\ref{fig6}(a)). Power spectral density analysis further revealed a shift in the dynamic signature. The $1/f$ self-energy spectrum for unrestrained G4 DNA was shown earlier in Fig.~\ref{fig4}(a). The imposition of restraints dramatically changes the behavior of $S(f)$ with the disappearance of fluctuations at low frequencies (see Fig.~\ref{fig6}(b)). Notice that the value of $\beta \simeq 0.66$ differs significantly from $\beta \simeq 0.90$ seen in Fig.~\ref{fig4}. The constrained DNA belongs to a different universality class than the free DNA mentioned earlier. This shows that inherent structural flexibility and large-amplitude conformational fluctuations of non-canonical DNA structures are the drivers of long-time temporal correlations and low-frequency $1/f$ noise in the internal energy dynamics.

As stated above, $C_{ii}(t)$ contains information equivalent to $S_{ii}(f)$. Nevertheless, it is relevant to make some important observations about $C_{ii}(t)$. In principle, the $1/f^\beta$ noise in $S_{ii}(f)$ (with $\beta < 1$) implies that there must be a similar power-law decay (scale free) in $C_{ii}(t) \sim t^{-(1-\beta)}$. Unfortunately, a simple log-log plot of $C_{ii}(t)$ vs. $t$ does not yield this power-law in a straightforward manner. This is a consequence of transients at early times and a high sampling variance at large $t$.

In conclusion, our results reveal an appealing universality in the nature and origins of $1/f$ noise in solvated DNA. In the presence of water and ions, different types of biologically important DNA structures show robust $1/f$ noise with universal exponent $\beta \simeq 0.90$. The only significant difference between G4 DNA, Hairpin DNA and Duplex DNA is the frequency range (or time range) over which dynamical modes are active, with significantly lower frequencies being excited for G4 DNA and Hairpin DNA. This is a consequence of the lower structural rigidity of these two forms of DNA. Our detailed simulations provide mechanistic insight into the physical basis of anomalous dynamics in various forms of DNA in solution, indicating that $1/f$ noise arises from a complex interplay between universal environmental coupling and intrinsic structure-specific fluctuations. We urge experimentalists to undertake detailed studies of these dynamical phenomena in order to provide further direction to simulations.

This refined understanding has broad implications for linking the DNA conformation, its interactions with the aqueous--ionic environment, and its dynamic behavior. Such insights are critical for elucidating the multifaceted biological roles of DNA, including gene regulation, telomere maintenance, and interactions with drugs and proteins, and open avenues for targeted strategies that harness dynamic control of DNA.

\newpage
\putbib

\newpage

\end{bibunit}
\clearpage
\setcounter{equation}{0}
\setcounter{figure}{0}
\setcounter{table}{0}
\begin{center}
\textbf{\large Supplementary Material: Universal $1/f$ Noise in the Power Spectra of Energy Time-series in Solvated DNA Dynamics}
\end{center}

\begin{bibunit}

\subsection{Molecular Dynamics Simulations}

Equilibrium MD simulations, each spanning 1 $\mu$s, were performed on three different DNA structures (G-quadruplex, Hairpin, Duplex). These simulations were carried out using the AMBER-18 software package \cite{Case2018}. The atomic coordinates for the initiation of MD simulations of the different DNA structures were acquired from the protein data bank (PDB). The PDB IDs are given in Table~\ref{tab1}. After taking the initial PDB configuration, the system was solvated with water and ions through the LEAP module of AMBER-18. The OL15 force field of DNA \cite{Zgarbov2015}, SPC/E for the water model, and the Jung-Cheatham parameters for the ions were used to perform the MD simulations. Subsequently, all three structures were charge-neutralized, and three independent MD simulation trajectories of 1 $\mu$s were performed, containing 100 mM NaCl or KCl concentrations as supplementary salt to maintain physiological conditions. Details about the total number of atoms, water, and ions are presented in Table~\ref{tab1} for each system.
\begin{table}[b]
\centering
\begin{tabular}{|l|c|c|c|}
\hline
{\bf Structure} & {\bf Total number of atoms} & {\bf Number of water molecules} & {\bf Number of ions} \\
\hline
G-quadruplex DNA & 18137 & 5790 & 35 K$^+$ 14 Cl$^-$ \\
{\it PDB Id: 1XAV} \cite{gquad.1xav} & & & \\
\hline
Hairpin DNA & 8332 & 2650 & 17 Na$^+$ 7 Cl$^-$ \\ 
{\it PDB Id: 5M1W} \cite{hairpin.5m1w} & & & \\
\hline
Duplex DNA & 26926 & 8702 & 42 Na$^+$ 20 Cl$^-$ \\ 
{\it PDB Id: 1D30} \cite{duplex.1d30} & & & \\
\hline
\end{tabular}
\caption{Structure and PDB ID of DNA Structure, Number of atoms, water molecules and ions in each system.}
\label{tab1}
\end{table}

The PDB structure of Duplex DNA contained the minor groove bound ligand DAPI. The Duplex DNA structure was prepared after DAPI was removed in the LEAP module and solvated with water and ions before simulation. For G4 DNA, the sequence 5'-TGAGGGTGGGGAGGGTGGGGAA-3' of the wild-type mycPu22 of NHE III of the c-MYC promoter oncogene was used \cite{Pu22}, and two thymine residues at positions 11 and 20 were replaced with guanines in the PDB structure of G4 DNA using the LEAP module of AMBER-18. In addition, two central K$^{+}$ ions are added between the three G-stacks of the quadruplex.

Before starting the MD simulations, we needed to eliminate unfavorable steric interactions without introducing artificial structural distortions. For this, the system underwent a relaxation procedure involving energy minimization, employing both {\it steepest descent} and {\it conjugate gradient} algorithms successively. In the first stage, DNA structures and ions were restrained with a force-constant of 25 kcal mol$^{-1}$ $\mbox{Å}^{-2}$. The water molecules were allowed to move freely for 1000 time steps while protecting the nucleic acid topology. Then, to remove conformational strain in DNA, the restriction on DNA was gradually reduced from 25 to 5 kcal mol$^{-1} \mbox{Å}^{-2}$ in five stages of 1000 time steps each. Following this, a restraint-free minimization was performed for 1000 steps. After the initial minimization, which successfully resolves severe steric clashes, the system is effectively frozen at 0 K.

This was followed by an equilibration period in which the MD simulation in the NVT ensemble was run for 20 ps. In this step, the system temperature was increased from 0 K to 300 K by the Langevin thermostat \cite{Uberuaga2004}, again with a restriction of 25 kcal mol$^{-1} $$\mbox{Å}^{-2}$ on DNA structures to protect nucleic acid topology. Then, to allow the system to adjust to its natural density at the target temperature, equilibration within the NPT ensemble was performed. This equilibration was executed in five discrete stages, each of 20 ps duration, during which the harmonic restraint applied to DNA was progressively reduced from 25 to 5 kcal mol$^-$$^1$ Å$^-$$^2$ in each step. Following this, a restraint-free 50 ps equilibration was performed within the NPT ensemble. To lock in the equilibrated density, the dimensions of the simulation box were fixed to the average volume observed during the last 50 ps of the NPT run. The system was subsequently subjected to 150 ps of NVT dynamics, after which the velocities of the final snapshot were rescaled to 300 K. Subsequently, before moving to the production run in the NPT ensemble, a 600 ps and 1 ns equilibration phase was conducted within the NPT ensemble; these simulations were performed using the SANDER module within the AMBER-18 software package. This was followed by an extended 5 ns equilibration period, also within the NPT ensemble in the PMEMD module of AMBER-18.

The above protocol ensures that the system is well equilibrated. Then, an all-atom production simulation of 1 $\mu$s duration was conducted for each of the three systems. These simulations were performed using the PMEMD module within the AMBER-18 software package, employing a 2 fs time-step. Data analysis was performed on the resulting stable trajectories. The SHAKE algorithm \cite{Ryckaert1977} was used to constrain bonds that involve hydrogen atoms. Long-range electrostatic interactions were treated using the Particle-Mesh Ewald (PME) summation method \cite{Darden1993}, with a non-bonded cutoff of 10 Å in real space. Coordinate data were recorded every 10 ps throughout the entire 1 $\mu$s trajectory for all systems. In addition to the above three trajectories, to check the effect of rigidity, an additional 500 ns MD run was conducted on the G-quadruplex DNA (GqDNA) structure with a restraint of 0.1 kcal mol$^{-1}$ Å$^{-2}$ during the entire production run.

The time series of various physical quantities $X(t)$ (electrostatic interaction energies in this letter) are used to compute the {\it autocorrelation function}
\begin{eqnarray}
C(t) &=& \langle X(\tau) X(\tau+t) \rangle - \langle X(\tau) \rangle \langle X(\tau+t) \rangle = \langle \delta X(\tau) \delta X(\tau+t) \rangle ,
\end{eqnarray}
where the angular brackets denote the translational averaging over $\tau$. In the stationary regime, which is of interest to us, this quantity is independent of the starting time $\tau$. The {\it power spectral density}, which is the Fourier transform of $C(t)$, is calculated using the Igor Pro software:
\begin{equation}
S(f_k) = \sum_{n=0}^{N-1} C(t_n) e^{-2\pi i k n/N} .
\end{equation}
Here, $S(f_k)$ is the complex value of the transform in the $k^{\rm th}$ frequency bin, C($t_n$) is the value of the correlation function at the $n^{\rm th}$ time step, $N$ is the total number of points in the wave.

\subsection{Power Spectra of Duplex DNA for Extended Frequency Range}

The power spectra in the main paper (Fig.~4) are computed by Fourier transforming the correlation function $C(t)$ in a time window ranging from $t=0$ (where $C(0)=1$) to $t=t_m$ where $C(t_m) \geq 10^{-3}$. For $t > t_m$, the correlation function fluctuates randomly around $C(t)=0$.

\begin{figure}[htb]
\includegraphics[width=0.7\textwidth]{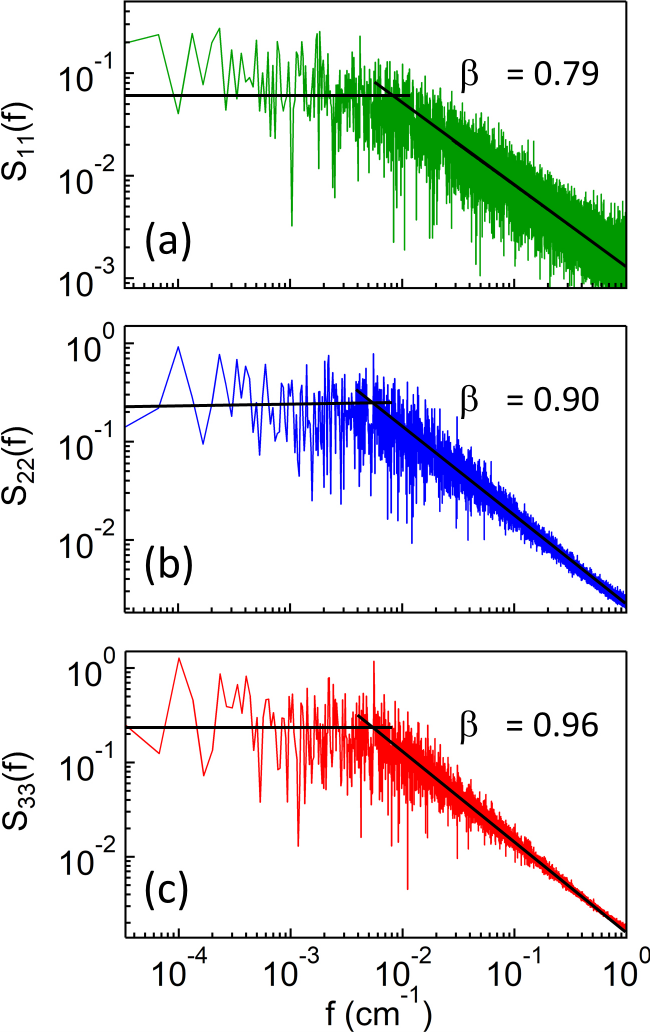}
\caption{Log-log plot of power spectra for Duplex DNA, $S_{ii}(f)$ vs. $f$, of energy time-series for $\delta E_1$, $\delta E_2$, $\delta E_3$. The solid lines show best linear fits with the power-law exponent $\beta$. The $S_{ii}(f)$ show white noise ($\beta = 0$) in the low-frequency regime, showing the absence of long-time correlations of structural fluctuations in the rigid Duplex DNA. We have confirmed that the data in the intermediate and high frequency regime is in agreement with that in Fig.~4 of the main paper. We see larger fluctuations in these power spectra because of the higher number of time points in the Fourier transform.}
\label{figS1}
\end{figure}

In Fig.~\ref{figS1}, we show the power spectra obtained for Duplex DNA by Fourier transformation of $C(t)$ in the complete time interval of our simulation, from $t=0$ to $t=1000$ ns.

\newpage
\putbib
\end{bibunit}

\end{document}